\documentclass[a4paper,conference]{IEEEtran}

\usepackage[nolist]{acronym}
\usepackage[cmex10]{amsmath}
\usepackage{amsfonts}
\usepackage{amssymb}
\usepackage{amssymb}
\usepackage{listings}
\usepackage{graphicx}
\usepackage{booktabs}

\begin{document}

\begin{acronym}
    \acro{ABI}{Application Binary Interface}
    \acro{AES-NI}{Advanced Encryption Standard -- New Instructions}
    \acro{AVX}{Advanced Vector Extensions}
    \acro{BPU}{Branch Prediction Unit}
    \acro{CCS}{Camera Coordinate System}
    \acro{CL}{Cache Line}
    \acro{CRI}{Core-Ring Interconnect}
    \acro{CT}{Computed Tomography}
    \acro{CUDA}{Compute Unified Device Architecture}
    \acro{DCS}{Detector Coordinate System}
    \acro{DP}{Double Precision}
    \acro{FBP}{Filtered Back Projection}
    \acro{FDK}{Feldkamp-Davis-Kress}
    \acro{GUp/s}{Giga Voxel Updates per Second}
    \acro{IACA}{Intel Architecture Code Analyzer}
    \acro{ICS}{Image Coordinate System}
    \acro{IF}{Instruction Fetcher}
    \acro{ILP}{Instruction Level Parallelism}
    \acro{IMCI}{Initial Many Core Instructions}
    \acro{ISPC}{Intel SPMD Program Compiler}
    \acro{L1}{Level One}
    \acro{L2}{Level Two}
    \acro{L3}{Level Three}
    \acro{LCPs}{Length Changing Prefixes}
    \acro{LLC}{Last Level Cache}
    \acro{MIC}{Many Integrated Cores}
    \acro{MPSS}{Manycore Platform Software Stack}
    \acro{MSE}{Mean Squared Error}
    \acro{MSR}{Model Specific Registers}
    \acro{NUMA}{Non-Uniform Memory Access}
    \acro{OpenCL}{Open Computing Language}
    \acro{PF}{Picker Function}
    \acro{PSNR}{Peak Signal-to-Noise Ratio}
    \acro{QPI}{QuickPath Interconnect}
    \acro{SHA}{Secure Hashing Algorithm}
    \acro{SIMD}{Single Instruction, Multiple Data}
    \acro{SM}{Streaming Multiprocessor}
    \acro{SP}{Single Precision}
    \acro{SSE}{Streaming SIMD Extensions}
    \acro{SVML}{Short Vector Math Library}
    \acro{TD}{Tag Directory}
    \acro{TD}{Tag Directory}
    \acro{TFlop/s}{Tera Floating-Point Operations per Second}
    \acro{TTS}{Time-to-Solution}
    \acro{VCS}{Voxel Coordinate System}
    \acro{VIPT}{Virtually Indexed, Physically Tagged}
    \acro{VPU}{Vector Processing Unit}
    \acro{WCS}{World Coordinate System}
\end{acronym}

\title{Performance Engineering for a Medical Imaging Application on the Intel Xeon Phi Accelerator}

\author{\IEEEauthorblockN{Johannes Hofmann}
\IEEEauthorblockA{Chair of Computer Architecture\\
University Erlangen--Nuremberg\\
Email: johannes.hofmann@fau.de}
\and
\IEEEauthorblockN{Jan Treibig, Georg Hager, Gerhard Wellein}
\IEEEauthorblockA{Erlangen Regional Computing Center\\
University Erlangen--Nuremberg\\
Email: jan.treibig@rrze.fau.de}
}

\maketitle

\begin{abstract}
    We examine the Xeon Phi, which is based on Intel's Many Integrated Cores
    architecture, for its suitability to run the FDK algorithm---the most
    commonly used algorithm to perform the 3D image reconstruction in cone-beam
    computed tomography. We study the challenges of efficiently parallelizing
    the application and means to enable sensible data sharing between threads
    despite the lack of a shared last level cache.  Apart from parallelization,
    SIMD vectorization is critical for good performance on the Xeon Phi; we
    perform various micro-benchmarks to investigate the platform's new set of
    vector instructions and put a special emphasis on the newly introduced
    vector gather capability. We refine a previous performance model for the
    application and adapt it for the Xeon Phi to validate the performance of
    our optimized hand-written assembly implementation, as well as the
    performance of several different auto-vectorization approaches.

 \end{abstract}

\section{Introduction}

The computational effort of 3D image reconstruction in \ac{CT} has required
special purpose hardware for a long time. Systems such as custom-built
FPGA-systems \cite{siemensfpga} and GPUs \cite{ctwasfirst,thumper} are still
widely-used today, in particular in interventional settings, where radiologists
require a hard time constraint for reconstruction. However, recently it has
been shown that today even commodity CPUs are capable of performing the
reconstruction within the imposed time-constraint \cite{fastrabbit}.
In comparison to traditional CPUs the Xeon Phi accelerator, which focuses on
numerical applications, is expected to deliver higher performance using the
same programming models such as C, C++, and Fortran.  Intel first began
developing the many-core design (then codenamed Larrabee) back in
2006---initially as an alternative to existing graphics processors.  In 2010
the original concept was abandoned and the design was eventually re-targeted as
an accelerator card for numerical applications.  The Xeon Phi is the first
product based on this design and has been available since early 2013 with 60
cores, a new 512\,bit wide SIMD instruction set, and 8 GiB of main memory.
This paper studies the challenges of optimizing the \ac{FDK} algorithm for the
Intel Xeon Phi accelerator.  The fastest available CPU implementation from
Treibig \textit{et al.} \cite{fastrabbit} served as starting point for the Xeon
Phi implementation. To produce meaningful and comparable results all
measurements are performed using the RabbitCT benchmarking framework
\cite{rohkohl2009}.

The paper is structured as follows. Section~2 will give an overview of previous
work about the performance optimization of this algorithm. A short introduction
to computed tomography is given in Section~3.  Section~4 introduces the
RabbitCT benchmark and motivates its use for this study. Next we provide a
hardware description of the Xeon Phi accelerator together with the results of
various micro-benchmarks in Section~5. In Section~6 we give an overview of the
implementation and the optimizations employed for the accelerator card.
Section~7 contains a detailed performance model for our application on the Xeon
Phi.  The results of our performance engineering efforts are presented in
Section~8; for the sake of completeness we also present the results obtained
with compiler-generated code. Finally we compare our results with the fastest
published GPU implementation and give a conclusion in Section~9.

\section{Related Work}
\label{sec:relatedwork}
Due to its medical relevance, reconstruction in computed tomography is a
well-examined problem. As vendors for \ac{CT} devices are constantly on the lookout
for ways to speed up the reconstruction time, many computer architectures have
been evaluated over time. Initially products in this field used special
purpose hardware based on FPGA and DSP designs \cite{siemensfpga}.  The
Cell Broadband Engine, which at the time of its release provided unrivaled
memory bandwidth, was also subject to experimentation \cite{cell1,cell2}.
It is noteworthy that \ac{CT} reconstruction was among the first non-graphics
applications that were run graphics processors \cite{ctwasfirst}.

However, the use of varying data sets and reconstruction parameters limited the
comparability of all these implementations. In an attempt to remedy this
problem, the RabbitCT framework \cite{rohkohl2009} provides a standardized,
freely available \ac{CT} scan data set and a uniform benchmarking interface
that evaluates both reconstruction performance and accuracy.  Current entries
in the RabbitCT ranking worth mentioning include \textit{Thumper} by Zinsser
and Keck \cite{thumper}, a Kepler-based implementation which currently
dominates all other implementations, and \textit{fastrabbit} by Treibig
\textit{et al.} \cite{fastrabbit}, a highly optimized CPU-based implementation.

\section{Computed Tomography}

In diagnostic and interventional computed tomography an X-ray source and a
flat-panel detector positioned on opposing ends of a gantry move along a
defined trajectory---mostly a circle or helix---around the patient; along the
way X-ray images are taken at regular angular increments.  In general 3D image
reconstruction works by back projecting the information recorded in the
individual X-ray images (also called projection images) into a 3D volume, which
is made up of individual voxels (volume elements). In medical applications, the
volume almost always has an extent of $512^3$ voxels.  To obtain the intensity
value for a particular voxel of the volume from one of the recorded projection
images we forward project a ray originating from the X-ray source through the
isocenter of the voxel to the detector; the intensity value at the resulting
detector coordinates is then read from the recorded projection image and added
to the voxel. This process is performed for each voxel of the volume and all
recorded projection images, yielding the reconstructed 3D volume as the result.

\section{RabbitCT Benchmarking Framework}
\label{sec:rabbitct}

Comparing different optimized \ac{FDK} implementations found in the literature
with respect to their performance can be difficult, because of variations in
data acquisition and preprocessing, as well as different geometry conversions
and the use of proprietary data sets. The RabbitCT framework \cite{rohkohl2009}
was designed as an open platform that tries to remedy the previously mentioned
problems. It features a benchmarking interface, a prototype back projection
implementation, and a filtered, high resolution \ac{CT} dataset of a rabbit;
also included is a reference volume that is used to derive various image
quality measures.  The preprocessed dataset consists of 496~projection
images that were acquired using a commercial C-arm \ac{CT} system. Each
projection is 1248$\times$960 pixels wide and stores the X-ray intensity values
as single-precision floating-point numbers. In addition, each projection comes
with a projection matrix $A \in \mathbb{R}^{3\times4}$, which is used to
perform the forward projection.  The framework takes care of all
required steps to set up the benchmark, so the programmer can focus entirely on
the actual back projection implementation, which is provided as a module
(shared library) to the framework.

\lstset{
        basicstyle=\footnotesize\ttfamily,
        breaklines=true,
        language=C,
        basicstyle=\small\ttfamily,
        numbers=left,
        numberstyle=\tiny,
        frame=tb,
        columns=fullflexible,
        showstringspaces=false,
        numbersep=2pt,
        numbers=right
}

\begin{lstlisting}[caption={\textsc{Unoptimized Reference Back Projection
Implementation Processing a Single Projection Image.}},
    label=src:fdk,
    float=htpb,
    captionpos=b,
    belowcaptionskip=4pt,
    ]
// iterate over all voxels in the volume
for (z = 0; z < L; ++z) {
  for (y = 0; y < L; ++y) {
    for (x = 0; x < L; ++x) {
      // convert to WCS
      float wx = O+x*MM;
      float wy = O+y*MM;
      float wz = O+z*MM;
      // forward projection
      float u = wx*A[0]+wy*A[3]+wz*A[6]+A[9];
      float v = wx*A[1]+wy*A[4]+wz*A[7]+A[10];
      float w = wx*A[2]+wy*A[5]+wz*A[8]+A[11];
      // dehomogenize 
      float ix = u/w;
      float iy = v/w;
      // convert to integer
      int iix = (int)ix;
      int iiy = (int)iy;
      // calculate interpolation weights
      float scalex = ix-iix;
      float scaley = iy-iiy;
      // load values for biliean interpolation
      float valbl = 0.0f; float valbr = 0.0f;
      float valtr = 0.0f; float valtl = 0.0f;
      if (iiy >= 0 && iiy < width &&
          iix >= 0 && iix < height)
        valbl = I[iiy * width + iix];
      if (iiy >= 0 && iiy < width &&
          iix+1 >= 0 && iix+1 < height)
        valbr = I[iiy * width + iix + 1];
      if (iiy+1 >= 0 && iiy+1 < width &&
          iix >= 0 && iix < height)
        valtl = I[(iiy + 1) * width + iix];
      if (iiy+1 >= 0 && iiy+1 < width &&
          iix+1 >= 0 && iix+1 < height)
        valtr = I[(iiy + 1)* width + iix + 1];
      // perform bilinear interpolation
      float valb =(1-scalex)*valbl+scalex*valbr;
      float valt =(1-scalex)*valtl+scalex*valtr;
      float val = (1-scaley)*valb+scaley*valt;
      // add distance-weighted results to voxel
      VOL[z*L*L+y*L+x] += val/(w*w);
    } // x-loop
  } // y-loop
} // z-loop
\end{lstlisting}

A slightly compressed version of the unoptimized reference implementation
that comes with RabbitCT is shown in Listing~\ref{src:fdk}. This code is
called once for every projection image.  The three outer \texttt{for} loops
(lines 2--4) are used to iterate over all voxels in the volume; note that we
refer to the innermost \texttt{x}-loop, which updates one ``line'' of voxels
in the volume, as line update kernel. The loop variables \texttt{x},
\texttt{y}, and \texttt{z} are used to logically address all voxels in
memory.  To perform the forward projection these logical coordinates used for
addressing must first be converted to the \ac{WCS}, whose origin coincides
with the isocenter of the voxel volume; this conversion happens in lines
6--8.  The variables \texttt{O} and \texttt{MM} that are required to perform
this conversion are precalculated by the RabbitCT framework and made
available to the back projection implementation in a \texttt{struct} pointer
that is passed to the back projection function as a parameter.  After this
the forward projection is performed using the projection matrix $A$ in lines
10--12.  In order to transform the affine mapping that implements the forward
projection into a linear mapping homogeneous coordinates are used.  Thus the
detector coordinates are obtained in lines~14 and~15 by dehomogenization.

In the next step a bilinear interpolation is performed. In order to do so,
detector coordinates are converted from floating-point to integer type
(lines~17 and~18), because integral values are required for addressing the
projection image buffer \texttt{I}. The interpolation weights \texttt{scalex}
and \texttt{scaley} are calculated in lines~20 and~21.  The four values needed
for the bilinear interpolation are fetched from the buffer containing the
intensity values in lines 25--36. The \texttt{if} statements make sure, that
the detector coordinates lie inside of the projection image; for the case where
the ray doesn't hit the detector, i.e.  the coordinates lie outside the
projection image an intensity value of zero is assumed (lines~23 and~24). Note
that the two-dimensional projection image is linearized, which is why we need
the projection image width in the variable \texttt{width}---also made available
by the framework via the \texttt{struct} pointer passed to the function---to
correctly address data inside the buffer.  The actual bilinear interpolation is
performed in lines 38--40.

Before the result is written back into the volume (line~42), it is weighed
according to the inverse-square law.  The variable \texttt{w}, which holds the
homogeneous coordinate $w$, contains an approximation of the distance from
X-ray source to the voxel under consideration and can be used to perform the
weighting.


%

\section{Intel Xeon Phi}

\begin{figure}[htbp]
    \centering
  \includegraphics[clip=true,width=\linewidth]{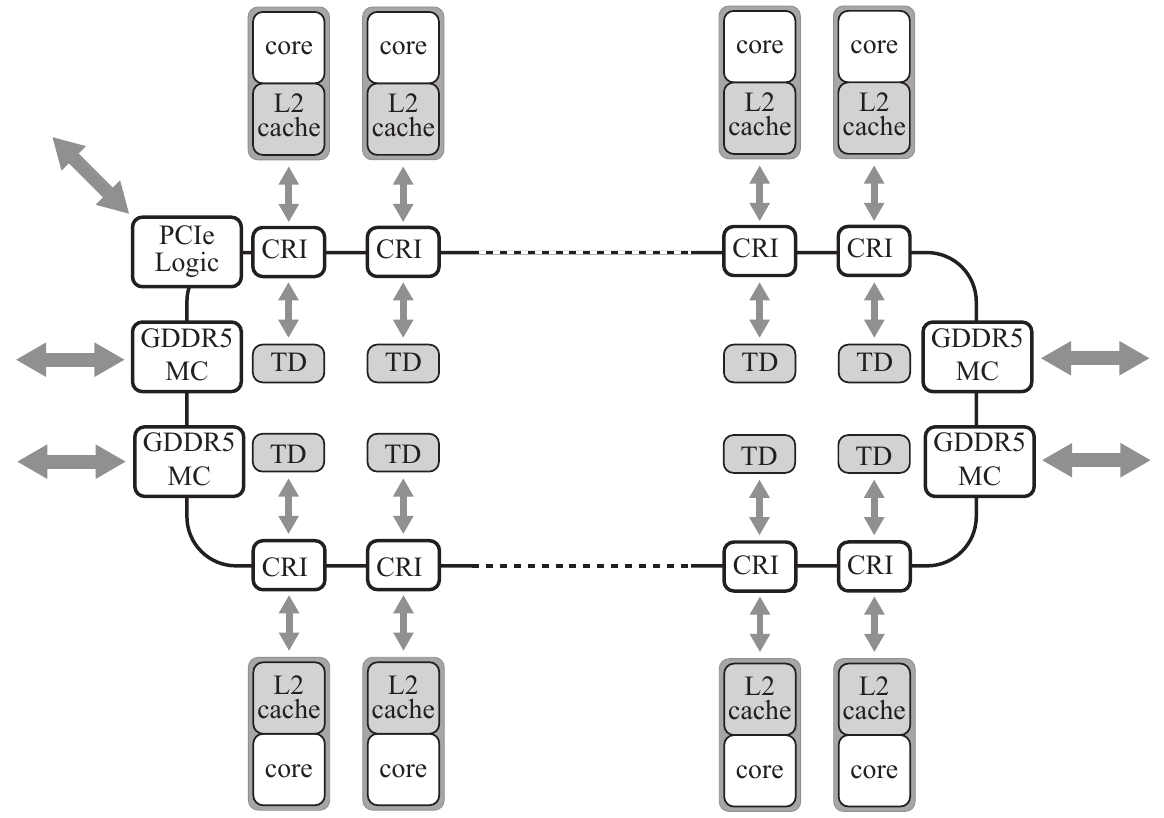}
  \caption{\textsc{Schematic Overview of the Xeon Phi 5110P Accelerator.}}
  \label{fig:mic-overview}
\end{figure}

An overview of the Xeon Phi 5110P is provided in Figure~\ref{fig:mic-overview}.
The main components making up the accelerator are the 60 cores connected to the
high bandwidth ring interconnect through their Core--Ring Interconnects (CRI);
interlaced with the ring is a total of eight memory controllers that connect
the processing cores to main memory as well as PCIe logic that communicates
with the host system.

The cores are based on a modified version of the P54C design used in the
original Pentium released in 1995.  Each core is clocked at 1.05~GHz and is a
fully functional, in-order core, which supports fetch and decode instructions
from four hardware thread execution contexts---twice the amount used in recent
x86 CPUs.  The superscalar cores feature a scalar pipeline (V-pipe) and a
vector pipeline (U-pipe). Connected to the U-pipe is the Vector Processing Unit
(VPU), which implements the new \ac{IMCI} vector extensions.

\subsection{Core Pipeline}

The cores used in the Xeon Phi are in-order, lacking all of the necessary logic
to manage out-of-order execution, making the individual cores less complex than
their traditional CPU counterparts.  A core can execute two instructions per
clock cycle: one on the V-pipe, which executes scalar instructions, prefetches,
loads, and stores; and one on the U-pipe, which can only execute vector
instructions.\footnote{Actual simultaneous execution is governed by a set of
non-trivial pairing rules \cite{pairing}.} The decode unit is shared by all
hardware contexts of a core and is a pipelined two-cycle unit to increase
throughput. This means it takes the unit two cycles to decode one instruction
bundle (i.e. one micro-op for the U- and one for the V-pipe); however, due to
its pipelined design the unit can deliver decoded bundles to \textit{different}
hardware threads each cycle.  As a consequence, at least two hardware threads
must be run on each core to achieve peak performance; using only one thread per
core will in the best case result in 50\% of peak performance. We found,
however, that it is good practise to always use all four hardware threads of a
core because most vector instructions have a latency of four clock cycles and
data hazards can be avoided without instruction reordering when using four
threads.

\subsection{Cache Organization, Core Interconnect, and Memory}

Most of Intel's cache concepts were adopted into the Xeon Phi: the \ac{CL} size
is 64\,bytes and cache coherency is is implemented across all caches using the
MESI protocol with the help of the distributed \ac{TD}.  Each core includes a
32\,KiB L1 instruction cache, a 32\,KiB L1 data cache, and a unified 512\,KiB
L2 cache.

The L1 cache is 8-way associative and has a 1 cycle latency for scalar loads
and a 3 cycle latency for vector loads. Its bandwidth has been increased to
64\,bytes per cycle, which corresponds exactly to the vector register width of
512\,bits.  In contrast to recent Intel x86 CPUs which contain two hardware
prefetching units for the L1 data cache (streaming prefetcher and stride
prefetcher), there exist no hardware prefetchers for the L1 cache on the Xeon
Phi. As a consequence, the compiler/programmer has to make heavy use of
software prefetching instructions---which are available in various flavors (cf.
Table~\ref{tab:prefetch})---to make sure data is present in the caches whenever
needed.

\begin{table}[htpb]
  \caption{Available Scalar Prefetch Instructions for the Intel Xeon Phi.}
  \begin{center}
    \begin{tabular}{ l c c c }
      \toprule
      Instruction & Cache Level & Non-temporal & Exclusive \\
      \midrule
      \texttt{vprefetchnta}     & L1    & Yes   & No \\
      \texttt{vprefetch0}       & L1    & No    & No \\
      \texttt{vprefetch1}       & L2    & No    & No \\
      \texttt{vprefetch2}       & L2    & Yes   & No \\
      \midrule
      \texttt{vprefetchenta}    & L1    & Yes   & Yes \\
      \texttt{vprefetche0}      & L1    & No    & Yes \\
      \texttt{vprefetche1}      & L2    & No    & Yes \\
      \texttt{vprefetche2}      & L2    & Yes   & Yes \\
      \bottomrule
    \end{tabular}
  \end{center}
  \label{tab:prefetch}
\end{table}

Apart from standard prefetches into the L1 and L2 caches (\texttt{vprefetch0,
vprefetch1}), there exist also variants that prefetch data into what Intel
refers to the L1/L2 non-temporal cache (\texttt{vprefetchnta, vprefetch2}).
Data prefetched into these non-temporal caches is fetched into the $n$th way
(associativity-wise) of the cache, where $n$ is the context id of the
prefetching hardware thread and made MRU---i.e. the most recently used data
will be replaced first.  Prefetches can also indicate the requested CL be
brought into the cache for writing, i.e.  in the exclusive state of the MESI
protocol (\texttt{vprefetche*}).

The L2 cache is 8-way associative and has a latency of 11 clock cycles. The
size of the L2 cache is twice the size of recent Intel x86 designs, namely
512\,KiB. The L2 cache contains a rudimentary streaming prefetcher that can
only detect strides up to 2\,CLs apart.

\begin{figure}[htbp]
    \centering
  \includegraphics[clip=true,width=\linewidth]{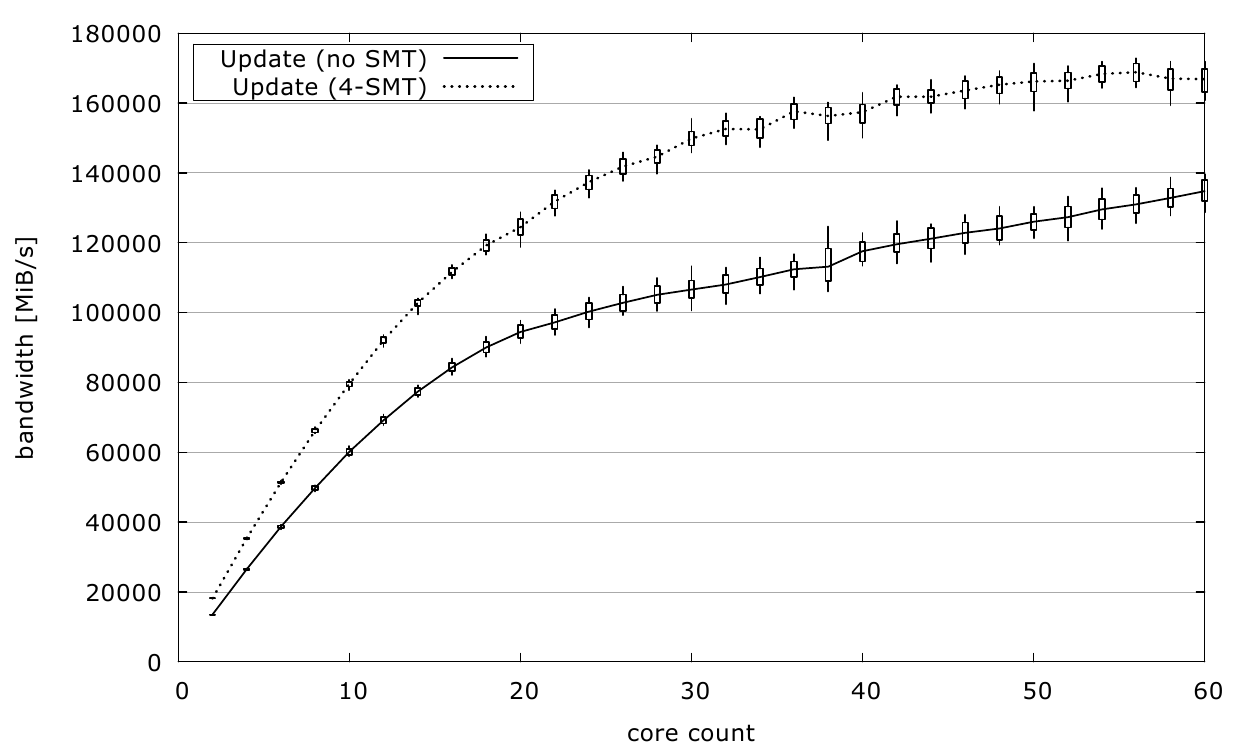}
  \caption{\textsc{Memory Bandwidth of Streaming Update Kernel.}}
  \label{fig:stream-update}
\end{figure}

The Xeon Phi contains a total of eight dual-channel GDDR5 memory controllers
clocked at 5\,GHz, yielding a theoretical peak memory bandwidth of 320\,GiB/s.
To get an estimate of the attainable bandwidth for our application, we ran a
streaming ``Update'' kernel which resembles the memory access pattern of our
application (cf.~Figure~\ref{fig:stream-update}).  We found that peak memory
performance can only be achieved by employing SMT. The bandwidth of about
165\,GiB/s corresponds to around 52\% of the theoretical peak performance; this
can be attributed to limited scalability of the memory system: the gradient of
the graph is steeper for e.g. 10--20 cores than it is for e.g.  50--60 cores.

\subsection{Initial Many Core Instructions}

While AVX2---the latest set of vector instructions for Intel x86
CPUs---provides a total of 16 vector registers, each 256\,bits wide, \ac{IMCI}
offers 32 register, each 512\,bits wide. \ac{IMCI} supports fused-multiply add
operations, yielding a maximum of 16 DP (32 SP) Flops per instruction.  In
addition to increasing the register count and width, \ac{IMCI} also introduces
eight vector mask registers, which can be used to mask out SIMD lanes in vector
instructions; this means that a vector operation is performed only selectively
on some of the elements in a vector register. Another novelty is the support
for vector scatter and gather operations.

\subsection{Vector Gather Operation}

In the \ac{FDK} algorithm, a lot of data has to be loaded from different
offsets inside the projection image.  The Intel Xeon Phi offers a vector gather
operation that enables filling of vector registers with scattered data. A major
advantage over sequential loads is the fact that vector registers can be used
for addressing the data; this means no detour of writing the contents of vector
registers to the stack to move them into scalar registers required for
sequential loads is necessary.

\lstdefinelanguage
   [x]{Assembler}
   [x86masm]{Assembler}
   {morekeywords={vmovaps, xor, vprefetchenta, vprefetche2,
   vgatherdps, jknzd, jkzd, vsubps, vmulps, vaddps}}
\lstset{
        breaklines=true,
        language=[x]Assembler,
        basicstyle=\small\ttfamily,
        numbers=left,
        numberstyle=\tiny,
        frame=htb,
        columns=fullflexible,
        showstringspaces=false,
        keepspaces=true,
        numbers=none
}
\begin{lstlisting}[caption={\textsc{Gather Primitive in Assembly. NB Particular Code Using Two Branches Shown Here Generated By C Intrinsic.}},
    label=src:gatherasm,
    float=htpb,
    captionpos=b,
    belowcaptionskip=4pt,
    ]
..L100: vgatherdps zmm6{k3}, [rdi+zmm13*4]
        jkzd       k3, ..L101
        vgatherdps zmm6{k3}, [rdi+zmm13*4]
        jknzd      k3, ..L100
..L101:
\end{lstlisting}

At first glance (cf. Listing~\ref{src:gatherasm}), a \texttt{vgatherdps}
instructions looks similar to a normal load instruction.  In the example
\texttt{zmm6} is the vector registers in which the gathered data will be
stored.  The \texttt{rdi} register contains a base address, \texttt{zmm13} is a
vector register holding 16~32\,bit integers which serve as offsets, and
\texttt{4} is the scaling factor. The 16 bits of the vector mask register act
as a write mask for the operation: if the $n$th bit is set to 1 the gather
instruction will fetch the data pointed at by the $n$th component of the
\texttt{zmm13} register and write it into the $n$th component of the
\texttt{zmm6} register; if the bit is set to 0 no data will be fetched and the
$n$th component of \texttt{zmm6} is not modified. When a gather instruction is
executed, only data from \textit{one} CL is fetched. This means that when the
data pointed at by the \texttt{zmm13} register is distributed over multiple CLs
the gather instruction has to be executed multiple times. To determine whether
all data has been fetched, the gather instruction will zero out the bits in the
vector mask register whenever the corresponding data was fetched. In
combination with the \texttt{jknzd} and \texttt{jkzd} instructions---which
perform conditional jumps depending of the contents of the vector mask
register---it is possible to form loop constructs to execute the gather
instruction as long as necessary to fetch all data, i.e.  until the vector mask
register contains all zero bits.

\begin{table}[htpb]
  \caption{Latencies in Clock Cycles of the Vector Gather Primitive.}
  \begin{center}
    \begin{tabular}{ c r r r r}
    \toprule
     Distribution &   \multicolumn{2}{c}{L1 Cache}  &  \multicolumn{2}{c}{L2 Cache} \\
                           \cmidrule(r){2-3}        \cmidrule(r){4-5}
                            & Instruction   & Loop  & Instruction   & Loop \\
      \midrule
        16 per CL           & 9.0           & 9.0   & 13.6          & 13.6 \\
        8 per CL            & 4.2           & 8.4   & 9.4           & 18.8 \\
        4 per CL            & 3.7           & 14.8  & 9.1           & 36.4 \\
        2 per CL            & 2.9           & 23.2  & 8.6           & 68.8 \\
        1 per CL            & 2.3           & 36.8  & 8.1           & 129.6\\
      \bottomrule
    \end{tabular}
  \end{center}
  \label{tab:gather}
\end{table}

A set of micro-benchmarks for likwid-bench from the likwid \cite{likwid-psti}
framework were devised to measure the cycles required to fetch data using
gather loop constructs; Table~\ref{tab:gather} shows the results, taking into
account distribution of data across CLs. We find that the latency for a single
gather instruction varies depending on how many elements it has to fetch from a
CL.  This might be taken as a hint that a single gather instruction itself is
implemented as yet another loop, this time in hardware---the larger the number
of elements that have to be fetched from a single CL, the higher the latency.

Table~\ref{tab:arch} summarizes the hardware specifications of the Xeon Phi and
integrates them with two state of the art reference systems from the CPU and
GPU domain.

\begin{table*}[tbp]
    \caption{Hardware Specifications of the Intel Xeon Phi and two State of the Art CPU and GPU Reference Systems.}
    \label{tab:arch}
    \centering
	\begin{tabular}{lccc}
	    \toprule
	    Microarchitecture           &IvyBridge-EP      &Knights Corner  & Kepler \\
	    Model                       &Xeon E5-2660 v2   &Xeon Phi 5110P  & Tesla~K20 (GK110)\\
	    \midrule
	    Clock                     &2.2\,GHz               &1.05\,GHz           &0.706\,GHz\\
        Sockets/Cores/Threads per Node  & 2/20/40          & 1/60/240       & 1/13/--\\
	    SIMD support                & 8\,SP/4\,DP    & 16\,SP/8\,DP & 192\,SP/64\,DP\\
        Peak TFlop/s                 & 0.70\,SP/0.35\,DP& 2.02\,SP/1.01\,DP & 3.52\,SP/1.17\,DP \\
	    \midrule
	    Node L1/L2/L3 cache        &20$\times$32\,KiB/20$\times$256\,KiB/20$\times$2.5\,MiB    &60$\times$32\,KiB/60$\times$512\,KiB/--  & 13$\times$48\,KiB + 13$\times$48\,KiB (read-only)/1.5\,MiB/--\\
	    Node Main Memory Configuration           &2$\times$4 ch. DDR3-1866   &16 ch. GDDR5 5\,GHz  & 10 ch. GDDR5 5.2\,GHz\\
        Node Peak Memory Bandwidth        &119.4\,GiB/s                &320\,GiB/s          & 208\,GiB/s     \\
	    \bottomrule
	\end{tabular}
\end{table*}

\section{Implementation}

Our implementation makes use of all the optimizations found in the original
\textit{fastrabbit} implementation \cite{fastrabbit}. As part of this work, we
improved the original clipping mask optimization\footnote{For some projection
angles several voxels are not projected onto the flat-panel detector. For these
voxels a zero intensity is assumed. Such voxels can be ``clipped'' off by
providing proper start and stop values for each \texttt{x}-loop.} by 10\%.
Another improvement we made was to pass function parameters inside vector
registers to the kernel in accordance with the Application Binary Interface
\cite{micabi}: instead of replicating values from scalar registers onto the
stack and then loading them into vector registers we directly pass the
parameters inside vector registers.

While register spilling was a problem in the original implementation, the Xeon
Phi with its 32 vector registers can handle all calculations without spilling.
The number arithmetic instructions can be greatly reduced by the use of the
fused multiply-add instructions (cf.  lines~6--8, 10--12, 38--40, and~42). All
divides are replaced with multiplications of the reciprocal; the reciprocal
instruction on the Xeon Phi provides higher accuracy than current CPU
implementations and is fully pipelined.

All projection data required for the bilinear interpolation are fetched using
gather loop constructs.  Several unsuccessful attempts to improve the L1 hit
rate of the gather instructions were made.  We found that the gather hint
instruction, \texttt{vgatherpf0hintdps}, is implemented as a dummy
operation---it has no effect whatsoever apart from instruction overhead.
Another prefetching instruction, \texttt{vgatherpf0dps}, appeared to be
implemented exactly the same as the actual gather instruction,
\texttt{vgatherdps}: instead of returning control back to the hardware context
after the instruction is executed, we found that control was relinquished only
\textit{after} the data has been fetched into the L1 cache, rendering the
instruction useless. Finally, scalar prefetching using the \texttt{vprefetch0}
instruction was evaluated. The problem with this approach is getting the $4
\cdot 16$~offsets stored inside a vector register into scalar registers. This
requires storing the contents of the vector register onto the stack and
sequentially loading them into general purpose registers. Obviously, 4~vector
stores, as well as 64 scalar loads and prefetches, amounting to a total of 132
scalar instructions, is too much instruction overhead. As a consequence we
evaluated variants in which only every second (68~instructions), fourth
(36~instructions), or eighth (20~instructions) component of the vector
registers was prefetched. Nevertheless, the overhead still outweighed any
benefits caused by increasing the L1 hit rate.

Because the application is instruction throughput limited, dealing with the if
statements (cf. lines 25--36 in Listing~\ref{src:fdk}) using the zero-padding
optimization\footnote{Zero-padding refers to an optimization involving
allocating a buffer that is large enough to ``catch'' all projection rays that
miss the detector; the original projection image is copied into the buffer and
the remainder of the buffer if filled with zero intensity values.  The
\texttt{if} statements to check whether the projected rays lie inside the
projection image are thus no longer necessary.} results in better performance
than the usage of predicated instructions, which incur additional instructions
to set the vector mask registers.

Despite the strictly sequential streaming access pattern inside the volume the
lack of a L1 hardware prefetcher mandates the use of software prefetching. We
also find that using software prefetching for the L2 cache results in a much
better performance than relying on the L2 hardware prefetcher. For the volume
data, we used prefetching with the exclusive hint, because the voxel data will
be updated.  In addition, we deliberately fetch the volume data into the
non-temporal portion of the L1 and L2 caches, because we know the volume is too
large\footnote{The volume memory footprint is $512^3$\,Voxels $\cdot$
4\,$\textrm{bytes}/\textrm{Voxel}$ = 512\,MiB.} to fit inside the caches; this
way, the volume data will not preempt cached projection data.  For prefetched
data be available when needed it is important to fetch the data in time. For
our application, we achieved best performance when prefetching volume data four
loop iterations before accessing them from main memory into the L2 cache and
one loop iteration ahead from the L2 into to L1 cache.

Efficient OpenMP parallelization requires more effort on the Xeon Phi than on
traditional CPUs.  While even on today's high-end multi-socket CPU systems the
number of hardware threads is usually below 100, the Xeon Phi features 240
hardware threads.  On CPUs it was sufficient to parallelize the outermost
\texttt{z}-loop (cf. line~2 in Listing~\ref{src:fdk}) and use a static
scheduling with chunk size of 1 to work around the imbalances created by the
clipping mask.  This way each thread is updating one plane of the volume a
time. On the Xeon Phi this distribution of work would result in 208 of the 240
threads updating two planes and 32 of the threads updating three planes. In
other words 208 threads would be idle 33\% of the time. The solution is to make
the amount of work more fine-granular, while at the same time ensuring the
amount of work will not become so small that the overall runtime is dominated
by overhead. To make the work more fine-granular the OpenMP collapse directive
was used to fuse the \texttt{z} and \texttt{y} loops. The optimum chunk size
was empirically determined to be 262---corresponding to about half a plane in
the volume.

Another important consideration on the Xeon Phi is thread placement.  The
default ``scatter'' thread placement, in which thread 0 is run on core 0,
thread 1 on core 1, etc. and SMT threads of cores are only used when all
physical cores have been exhausted proves unfit for our application.  With this
scattered placement threads that run on the same physical core have no spatial
locality in the volume; as a result they do not have a spatial locality in the
projection image, which leads to preemption of projection data in the core's
caches (which is shared among the hardware contexts of the core). Using
``gather'' thread placement in which thread 0 runs on hardware context 0 of
core 0, thread 1 on context 1 of core 0, etc. we ensure spatial locality inside
the volume and the projection data thus reducing cache preemptions.

\section{Performance Model}
Popular performance models like the Roofline model
\cite{Williams:EECS-2008-134} reduce investigations to determining whether
kernels are compute- or memory-bound, not taking runtime contributions of the
cache subsystem into account.

The performance model we use is based on a slightly modified version of the
model used in the original \textit{fastrabbit} publication
\cite{fastrabbit,th09}.  At the basis of the model is the execution time
required to update the 16 voxels in a single CL, assuming all data is available
in the L1 cache.  In addition, the contribution of the cache and memory
subsystem is modeled, which accounts for time spent transferring all data
required for the update into the L1 cache and back. In the original model,
designed for out-of-order CPUs, an estimation whether the cache subsystem
overhead can be hidden by overlapping it with the execution time is given and
the authors conclude that there exist sufficient suitable
instructions\footnote{Because the L1 cache is single-ported---i.e. it can only
communicate with either the core or the L2 cache at any given clock
cycle---transfers between the L1 and L2 caches can only overlap with
``suitable'' instructions that do not access the L1 cache such as, e.g.,
arithmetic instructions with register operands.} to hide any overhead caused by
in-cache transfers.  However, in their analysis, Treibig \textit{et al.} only
consider the in-cache contribution of the CLs relating to the voxel volume; all
CLs pertaining to the projection images, required for the bilinear
interpolation, are assumed to reside in the L1 cache. On the Intel Xeon Phi we
find this simplification no longer holds true. There is a non-negligible cost
for transferring the projection data from the L2 to the L1 cache that can not
be overlapped with the execution time.

\subsection{Core Execution Time}

Unfortunately there exist no tools such as, e.g., the Intel Architecture Code
Analyzer (IACA) \cite{iacaweb}, which is used to measure kernel execution times
on Intel's CPU microarchitectures, for the Intel Xeon Phi. Therefore, we have
to perform a manual estimation of the clock cycles spent in a single iteration
of the line update kernel---which corresponds to the update of one CL.  To
complicate things, simply counting the instructions in the kernel is not an
option, because the number of gather instructions varies depending on the
distribution of the data to be fetched across CLs. As a consequence, we begin
with an estimation of the execution time for a gather-less kernel (i.e. a
version of the line update kernel in which all gather loop constructs have been
commented out).

Manually counting the instructions, we arrive at 34~clock cycles for a
gather-less kernel iteration.  This analytical estimation was verified by
measurement.  For one voxel line containing 512 voxels a runtime of 2402~clock
cycles was measured using a single thread. This corresponds to 75~clock cycles
per kernel iteration (when one iteration updates 16 voxels). Taking into
account that the single thread can only issue instruction every other clock
cycle, the core execution time for one loop iteration is approximately
37.5~clock cycles---which is a close fit to the value of 34~clock cycles
determined previously. For our model, we use the measured value of 37.5~clock
cycles because it contains non-negligible overhead that was not accounted for
in the analytical value.\footnote{ The overhead includes the time it takes to
call the line update kernel (backing up and later restoring callee-save
registers, the stack base pointer, etc.  onto the stack) as well as
instructions in the kernel that are not part of the loop body, such as
resetting the loop counter.}

To estimate the contribution of the gather loop constructs we first determine
how often a gather instruction is executed on average for a CL update. To get
this value, we divide the total number of gather instructions issued during the
reconstruction (obtained by measurement) by total number of loop iterations. We
find that, on average, the gather instruction is executed 16 times in a kernel
iteration. Distributing that number over the four gather loop constructs (one
for each of the four values required for the bilinear interpolation) we arrive
at 4 gather instructions per gather loop---indicating that the data is, on
average, distributed across four CLs.  From this we can infer the runtime
contribution based on our previous findings (cf.  Table~\ref{tab:gather}). The
latency of each gather instruction in the situation where the data is
distributed across four CLs is 3.7~clock cycles.  With a total of 16~gather
instructions per iteration, the contribution is 59.2~clock cycles. Together
with the remaining part of one kernel loop iteration (37.5~clock cycles), the
total execution time is approximately 97~clock cycles.

\subsection{Cache and Memory Subsystem Contribution}

To estimate the impact of the runtime spent transferring the data required for
the CL update we first have to identify which transfers can not be overlapped
with execution time.  As previously established the voxel volume is too large
for the caches. Thus each CL of the volume has to be brought in from main
memory for the update; eventually, the updated CL will also have to be evicted.
This means that a total of 2\,CLs, corresponding to 128\,byte, have to be
transfered. Using software prefetching any latency and transferring cost from
the memory and cache subsystems regarding volume data can be avoided.

As previously discussed, prefetching the projection data is not possible
without serious performance penalties.  Using likwid-perfctr from the likwid
framework \cite{likwid-psti} we investigated the Xeon Phi's performance
counters and found that 88.5\% of the projection data can be serviced
from the local L1 cache and the remaining 11.5\% can be serviced from
the local L2 cache.  Since each gather is transferring a full CL, this
amounts to approximately $16\,\textrm{CLs} \cdot 64\,\textrm{byte/CL} \cdot
11.5\% \approx 118\,\textrm{byte}$. We estimate the \textit{effective} L2
bandwidth in conjunction with the gather instruction to be the following: the
latency of a single gather instruction (when dealing with data that is
distributed across four CLs) was previously measured to be 3.7\,clock cycles
with data in L1 cache, respectively 9.1\,clock cycles with data in the L2 cache
(cf.  Table~\ref{tab:gather}).  Assuming the difference of 5.4\,clock cycles to
be the exclusive L2 cache contribution, we arrive at an effective bandwidth of
$64\,\textrm{byte} / 5.4\,\textrm{cycle} = 11.85\,\textrm{byte/cycle}$. The
average memory subsystem contribution is thus $118\,\textrm{byte} /
11.85\,\textrm{byte/cycle}\approx10\,\textrm{cycles}$.

\begin{figure}[htbp]
    \centering
  \includegraphics[clip=true,width=0.9\linewidth]{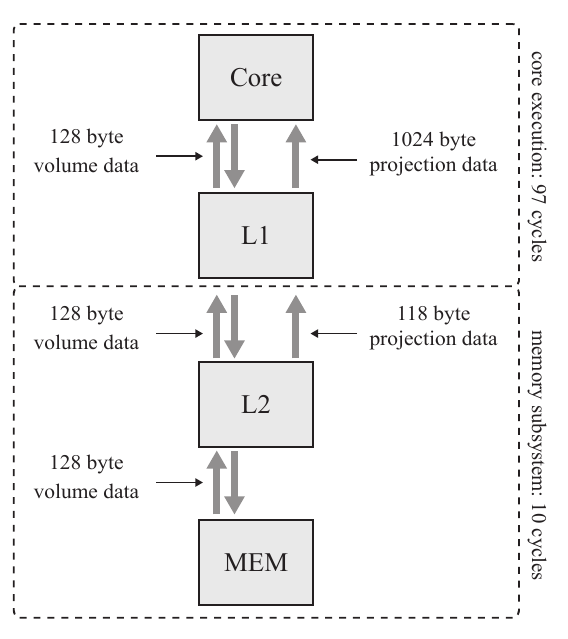}
  \caption{\textsc{Overview of Execution Time and Memory Subsystem
Contribution.}}
  \label{fig:ecm-cycles}
\end{figure}

Figure~\ref{fig:ecm-cycles} provides an overview of the performance model.  The
upper part shows core and L1 cache, together with all data transfers from the
cache. The lower part shows the memory hierarchy through which data has to be
transfered to perform the CL update.  The arrows to the left represent the CLs
pertaining the voxel volume data; prefetching these CLs in time guarantees
overlap of transfers with core execution. The arrow to the right between the L1
and L2 caches represents the transfers of projection data which can not be
prefetched; the latency of these transfers is the determining factor for the
memory subsystem contribution.  This leads to a total of 107\,clock cycles to
perform a single CL update.

Based on the runtime of a single kernel iteration we can determine whether the
memory bandwidth becomes a limiting factor for our application. For each loop
iteration, 128\,byte (2\,CLs) have to be transfered over the memory interfaces.
Each of the 60 cores is clocked at 1.05\,GHz; at 107\,cycles per iteration, the
required bandwidth is:
\[ \frac{1.05\,\textrm{GHz/core}}{107\,\textrm{cycles}} \cdot
60\,\textrm{cores} \cdot 128\,\textrm{byte} = 70.0\,\textrm{GiB/s.} \]

The required value is well below the measured sustainable bandwidth of around
165\,GiB/s (cf. Fig.~\ref{fig:stream-update}), indicating that bandwidth is not a
problem for our application.

Given the model, the total runtime contribution of the line update kernel is \[
\frac{4.39\cdot 10^{10}\,\textrm{voxels}}{16\,\textrm{voxels/iteration}} \cdot
\frac{107\,\textrm{cycles/iteration}}{60\,\textrm{cores} \cdot
1.048\,\textrm{GHz/core}} = 4.67\,\textrm{s.}\footnote{The total number of
voxels to process is given by considdering each voxel of the clipped volume once
for each of the 496~projection images.} \]

Unfortunately, there is a non-negligible amount of time spent outside of the
line update kernel. The value obtained by measuring the runtime of the
reconstruction with the call to the kernel commented out was 0.42\,s. Thus, the
total reconstruction time is 5.09\,s.  Foreclosing the runtime of the assembly
implementation from the next section which is 5.16 seconds (cf.
Table~\ref{tab:results}) we estimate the model error at 1.4\%.

\section{Results and Discussion}
\label{sec:results}

In addition to our hand-written assembly implementation we also evaluated
several auto-vectorization approaches for the FDK kernel: native vectorization
using the Intel C Compiler, the only recently introduced vectorization
directive from the latest OpenMP 4~standard \cite{omp4} implemented in the
Intel Compiler, and the \ac{ISPC} \cite{ispc}.  Table~\ref{tab:results} shows
the runtime in seconds and the corresponding performance in \ac{GUp/s}---the
commonly used performance metric for \ac{FDK}---of all implementations.  All
implementations were benchmarked using static OpenMP scheduling with a chunk
size of 262 voxel lines; independent of the implementation, this value resulted
in the best performance.

\begin{table}[tpb]
  \caption{Runtimes and Performance of All Implementations for a $512^3$
Volume.}
  \begin{center}
    \begin{tabular}{ l c c }
      \toprule
      Version & time [s] & Performance [GUp/s]         \\
      \midrule
      OpenMP 4 (\texttt{\#pragma simd})              & 7.77 & 5.6 \\
      ISPC (Version 1.5.0)                               & 7.00 & 6.3 \\
      Intel C Compiler (Version 13.1.3)                  & 6.99 & 6.3 \\
      Assembly                                      & 5.16 & 8.5 \\
      \bottomrule
    \end{tabular}
  \end{center}
  \label{tab:results}
\end{table}

We find that the performance of auto-vectorization variants can not match the
speed of our manually written assembly kernel. Even the best of the three
variants---the native vectorization of the Intel C Compiler---can only provide
around 74\% of the performance of hand-written code. We find that the
performance provided by the latest version of the free, open-source \ac{ISPC}
almost matches that of the commercial Intel C Compiler; the original \ac{ISPC}
version that was used during the early stages of this work had a 10\% lower
performance. The result obtained with the OpenMP~4 directive was the worst; the
main reason is that the standard guarantees the results obtained with this
vectorization are identical to that of scalar code---thus prohibiting several
optimizations, such as reordering of arithmetic instructions to increase
performance.

Our performance model revealed that even in the ideal case in which all
projection data resides in the L1 cache, the runtime impact of gathering the
projection data (59.2~cycles) dominates the overall runtime of a kernel
iteration (97~cycles).  We thus identify the gather operation as the limiting
factor for this application.  While all other parts of the \ac{FDK} kernel
benefit from the increase of the vector register width at the same time the
increased width counteracts the performance, because the cost of filling the
vector registers with scattered data increases linearly with register width.

\section{Conclusion}

We have presented a detailed examination of the Intel Xeon Phi accelerator by
performing various benchmarks. We performed various optimizations for the
\ac{FDK} algorithm and devised a manually vectorized assembly implementation
for the Xeon Phi and compared it to auto-vectorized code.  In order to
integrate our findings with today's state of the art reconstruction
implementations a comparison of our implementations with an
improved\footnote{Back-porting various optimizations of the Xeon Phi
implementations yielded a 25\% increase in performance for the
\textit{fastrabbit} CPU implementation.} version of the \textit{fastrabbit}
implementation, as well as the fastest currently available GPU implementation
called \textit{Thumper} \cite{thumper} is shown in
Table~\ref{tab:7:results_gups}.

\begin{table}[htpb]
  \caption{Comparison of Different Platforms in GUp/s.}
  \begin{center}
    \begin{tabular}{ l l c c }
      \toprule
      Platform & Version & $512^3$ & $1024^3$ \\
      \midrule
      2S-Xeon E5-2660 & \textit{improved fastrabbit}         & 6.2 & 6.7 \\
      Xeon Phi 5110P & OpenMP 4 (\texttt{\#pragma simd}) & 5.6 & 5.7  \\
      &ISPC (Version 1.5.0)                             & 6.3 & 6.4 \\
      &Intel C Compiler (Version 13.1.3)                                   & 6.3 & 6.4 \\
      &Assembly                                              & 8.5 & 13.1 \\
      GeForce GTX\,680 & \textit{Thumper} & 67.7 & 88.2 \\
      \bottomrule
    \end{tabular}
  \end{center}
  \label{tab:7:results_gups}
\end{table}

We find that the Kepler-based GeForce GTX~680 by Nvidia can perform the
reconstruction 7--8 times faster, depending on the volume's discretization.
This discrepancy can not be explained by simply examining the platforms'
specifications such as peak Flop/s and memory bandwidth.  The main causes
contributing to the GPU's superior performance for this particular application
are discussed in the following.

Most computations involved in the reconstruction kernel, such as the projection
of voxels onto the detector panel or the bilinear interpolation, are typical
for graphics applications (which GPUs are designed for). While, due to the
fused multiply-add operation, the forward projection is performed efficiently
on both the GPU and the Xeon Phi platform, the bilinear interpolation is not.
GPUs posses additional hardware called texture units, each of which can perform
a bilinear interpolations using a single instruction for data inside the
texture cache.  To emphasize the implications, consider that out of the total
of 97~clock cycles for one loop iteration of the \ac{FDK} kernel, 6~cycles are
used for the computation of the detector coordinates and 3~cycles to weight the
interpolated intensity value and update the voxel volume; the remaining
88\,clock cycles, more than 90\% of the kernel, is spent on the bilinear
interpolation\footnote{This includes preparing the interpolation weights,
converting floating-point detector coordinates to integral values, gathering
the projection data, and performing the actual interpolation.}---which is
handled by a single instruction on a GPU.

Given a sufficient amount of work, Nvidia's CUDA programming model does a
better job at hiding latencies. As seen before, even in the ideal case where
all data can be serviced from the L1 cache, on average, each of the gather
instructions has a latency of 3.7\,clock cycles.  Although the Intel Xeon Phi
can hide the latencies of most instructions when using all four hardware
contexts of a core, 4-way SMT is not sufficient to hide latencies caused by
loading non-continuous data.  In contrast to SMT, Nvidia's multiprocessors
feature hardware that allows them to instantly switch between
warps.\footnote{On Nvidia GPUs, the number of CUDA threads concurrently
executing on a core is called warp.} This way, every time a warp has to wait
for an instruction to complete or data to arrive from the caches or main
memory, the hardware simply schedules another warp in the meantime.  Given a
sufficient number of warps to choose from, this approach can hide much higher
latencies than the 4-way SMT in-order approach.

Although we have shown that the Intel Xeon Phi accelerator can not provide the
same performance as GPUs for the task of 3D reconstruction in the
interventional setting, there nevertheless might be applications that can
benefit from our work.  One promising application seems to be the
reconstruction of large \ac{CT} volumes.  Today, the largest industrial \ac{CT}
scanner, which at the time of this writing is the XXL-CT device only recently
installed by the Fraunhofer Institute in F\"urth \cite{xxlct}, is capable of
recording projection images with a resolution of 10000$\times$10000\,pixels,
corresponding to more than 380\,MiB per projection image. In this setting, it
is possible for main memory capacity and bandwidth to play more important
roles, potentially giving CPUs, with their high memory capacities, and the
Intel Xeon Phi, with its high memory bandwidth,  an advantage over GPUs.
Another interesting topic of research, of course, will be to evaluate the next
iteration of the Intel Xeon Phi architecture, codenamed Knights Landing, for
this application once it becomes available.

\bibliographystyle{IEEEtran}
\bibliography{rrze}

\begin{thebibliography}{10}
\providecommand{\url}[1]{#1}
\csname url@samestyle\endcsname
\providecommand{\newblock}{\relax}
\providecommand{\bibinfo}[2]{#2}
\providecommand{\BIBentrySTDinterwordspacing}{\spaceskip=0pt\relax}
\providecommand{\BIBentryALTinterwordstretchfactor}{4}
\providecommand{\BIBentryALTinterwordspacing}{\spaceskip=\fontdimen2\font plus
\BIBentryALTinterwordstretchfactor\fontdimen3\font minus
  \fontdimen4\font\relax}
\providecommand{\BIBforeignlanguage}[2]{{%
\expandafter\ifx\csname l@#1\endcsname\relax
\typeout{** WARNING: IEEEtran.bst: No hyphenation pattern has been}%
\typeout{** loaded for the language `#1'. Using the pattern for}%
\typeout{** the default language instead.}%
\else
\language=\csname l@#1\endcsname
\fi
#2}}
\providecommand{\BIBdecl}{\relax}
\BIBdecl

\bibitem{siemensfpga}
B.~Heigl and M.~Kowarschik, ``{High-speed reconstruction for C-arm computed
  tomography},'' in \emph{{In Proceedings Fully 3D Meeting and HPIR Workshop}},
  July 2007, pp. 25--28.

\bibitem{ctwasfirst}
\BIBentryALTinterwordspacing
G.~Pratx and L.~Xing, ``Gpu computing in medical physics: A review,''
  \emph{Medical Physics}, vol.~38, no.~5, pp. 2685--2697, 2011. [Online].
  Available: \url{http://link.aip.org/link/?MPH/38/2685/1}
\BIBentrySTDinterwordspacing

\bibitem{thumper}
T.~Zinsser and B.~Keck, ``{Systematic Performance Optimization of Cone-Beam
  Back-Projection on the Kepler Architecture},'' in \emph{Proceedings of the
  12th Fully Three-Dimensional Image Reconstruction in Radiology and Nuclear
  Medicine}, F.~committee, Ed., 2013, p. 225–228.

\bibitem{fastrabbit}
\BIBentryALTinterwordspacing
J.~Treibig, G.~Hager, H.~G. Hofmann, J.~Hornegger, and G.~Wellein, ``Pushing
  the limits for medical image reconstruction on recent standard multicore
  processors,'' \emph{International Journal of High Performance Computing
  Applications}, 2012, (Accepted). [Online]. Available:
  \url{http://arxiv.org/abs/1104.5243}
\BIBentrySTDinterwordspacing

\bibitem{rohkohl2009}
C.~Rohkohl, B.~Keck, H.~Hofmann, and J.~Hornegger, ``{RabbitCT - an open
  platform for benchmarking 3D cone-beam reconstruction algorithms},''
  \emph{Medical Physics}, vol.~36, no.~9, pp. 3940--3944, 2009.

\bibitem{cell1}
\BIBentryALTinterwordspacing
M.~Kachelriess, M.~Knaup, and O.~Bockenbach, ``Hyperfast parallel-beam and
  cone-beam backprojection using the cell general purpose hardware.'' \emph{Med
  Phys}, vol.~34, no.~4, pp. 1474--86, 2007. [Online]. Available:
  \url{Http://www.biomedsearch.com/nih/Hyperfast-parallel-beam-cone-backprojection/17500478.html}
\BIBentrySTDinterwordspacing

\bibitem{cell2}
H.~Scherl, M.~Kowarschik, H.~G. Hofmann, B.~Keck, and J.~Hornegger,
  ``Evaluation of state-of-the-art hardware architectures for fast cone-beam ct
  reconstruction,'' \emph{Parallel Comput.}, vol.~38, no.~3, pp. 111--124, Mar.
  2012.

\bibitem{pairing}
\BIBentryALTinterwordspacing
``{Intel Xeon Phi Coprocessor Vector Microarchitecture}.'' [Online]. Available:
  \url{{http://software.intel.com/sites/default/files/article/393199/intel-xeon-phi-coprocessor-vector-microarchitecture.pdf}}
\BIBentrySTDinterwordspacing

\bibitem{likwid-psti}
\BIBentryALTinterwordspacing
J.~Treibig, G.~Hager, and G.~Wellein, ``{LIKWID}: A lightweight
  performance-oriented tool suite for x86 multicore environments,'' in
  \emph{PSTI2010, the First International Workshop on Parallel Software Tools
  and Tool Infrastructures}.\hskip 1em plus 0.5em minus 0.4em\relax Los
  Alamitos, CA, USA: IEEE Computer Society, 2010, pp. 207--216. [Online].
  Available: \url{http://dx.doi.org/10.1109/ICPPW.2010.38}
\BIBentrySTDinterwordspacing

\bibitem{micabi}
{Intel Corporation}, \emph{{System V Application Binary Interface --- K1OM
  Architecture Processor Supplement}}, April 2012.

\bibitem{Williams:EECS-2008-134}
S.~W. Williams, A.~Waterman, and D.~A. Patterson, ``Roofline: An insightful
  visual performance model for floating-point programs and multicore
  architectures,'' EECS Department, University of California, Berkeley, Tech.
  Rep. UCB/EECS-2008-134, Oct 2008.

\bibitem{th09}
J.~Treibig and G.~Hager, ``Introducing a performance model for
  bandwidth-limited loop kernels,'' in \emph{Parallel Processing and Applied
  Mathematics}, ser. Lecture Notes in Computer Science, R.~Wyrzykowski,
  J.~Dongarra, K.~Karczewski, and J.~Wasniewski, Eds.\hskip 1em plus 0.5em
  minus 0.4em\relax Springer Berlin / Heidelberg, 2010, vol. 6067, pp.
  615--624.

\bibitem{iacaweb}
\BIBentryALTinterwordspacing
``Intel architecture code analyzer.'' [Online]. Available:
  \url{http://software.intel.com/en-us/articles/intel-architecture-code-analyzer/}
\BIBentrySTDinterwordspacing

\bibitem{omp4}
{OpenMP Architecture Review Board}, \emph{{OpenMP Application Program Interface
  --- Version 4.0}}, July 2013.

\bibitem{ispc}
M.~Pharr and W.~R. Mark, ``{ispc: A SPMD Compiler for High-Performance CPU
  Programming},'' in \emph{In Proceedings Innovative Parallel Computing
  (InPar)}, San Jose, CA, May 2012.

\bibitem{xxlct}
\BIBentryALTinterwordspacing
``{Fraunhofer Institute, XXL-CT}.'' [Online]. Available:
  \url{{http://www.iis.fraunhofer.de/de/bf/xrt/system/xxl-ct.html}}
\BIBentrySTDinterwordspacing

\end{thebibliography}
\end{document}